# Assessing Pedagogical Readiness for Digital Innovation: A Mixed-Methods Study

Ning Yulin and Danso Solomon Danquah

**Abstract:** Digital innovation in education has revolutionized teaching and learning processes, demanding a rethink of pedagogical competence among educators. This study evaluates the preparation of instructors to use digital technologies into their educational practices. The study used a mixed-methods approach, integrating both qualitative interviews and quantitative surveys to evaluate teachers' institutional support systems, beliefs, and technical proficiency. The results show that even while a large number of educators acknowledge the benefits of digital tools, problems including poor professional development and change aversion still exist. In order to improve digital pedagogical preparation, the study emphasizes the necessity of focused training initiatives and encouraging institutional regulations. There is discussion on the implications for educational institutions and policymakers.

## Introduction

**Background**

The rapid evolution of digital technologies has fundamentally transformed the educational landscape, necessitating a paradigm shift in pedagogical practices. Educators are now expected to be digitally competent, adaptable, and innovative in their teaching approaches to meet the demands of 21st-century learners (Tondeur et al., 2017). This shift is driven by several factors, including advancements in technology, changing student learning preferences, and the global push toward modernizing instructional methods (Koehler & Mishra, 2009). Digital innovation in education is no longer a luxury but a necessity, as it enhances learning outcomes, fosters student engagement, and prepares learners for a technology-driven world (Selwyn, 2016).

Pedagogical readiness [1], a critical component of successful technology integration, encompasses multiple dimensions. These include teachers' digital competence, their attitudes toward technology, and the institutional infrastructure that supports technology adoption (Ertmer et al., 2012). Digital competence refers to the ability to use digital tools effectively and ethically, while attitudes toward technology influence educators' willingness to embrace innovation (Redecker & Punie, 2017). Institutional infrastructure, on the other hand, includes access to resources, professional development opportunities, and policies that encourage technological engagement (Al-Furaih & Al-Awidi, 2018). Together, these dimensions form the foundation for effective digital pedagogy.

A key factor influencing pedagogical readiness is teachers' belief systems, which play a pivotal role in shaping their willingness to adopt and integrate new technologies (Scherer et al., 2019). Research indicates that teachers with positive attitudes toward technology are more likely to incorporate digital tools effectively in their classrooms, leading to improved student outcomes (Howard & Mozejko, 2015). However, the relationship between beliefs and practices is complex. While some educators enthusiastically embrace digital innovation, others remain hesitant due to a lack of confidence, fear of failure, or skepticism about the educational value of technology (Ertmer et al., 2012). These barriers are often compounded by contextual challenges, such as limited access to resources, inadequate training, and insufficient institutional support (Al-Furaih & Al-Awidi, 2018).

Professional development programs are essential for addressing these challenges and fostering a culture of innovation among educators. Effective training initiatives should focus not only on technical skills but also on pedagogical strategies that align with teachers' existing practices and beliefs (Cheah et al., 2023). For instance, the Technological Pedagogical Content Knowledge (TPACK) framework emphasizes the intersection of technology, pedagogy, and content knowledge, providing a structured approach to enhancing teachers' digital preparedness (Koehler & Mishra, 2009). Similarly, the PICRAT model offers a practical framework for evaluating how teachers use technology in their classrooms, encouraging them to move from passive to transformative practices (Kimmons et al., 2020).

Global initiatives, such as the European Framework for the Digital Competence of Educators (DigCompEdu), highlight the importance of preparing teachers for the digital era (Redecker & Punie, 2017). These frameworks provide guidelines for developing digital competencies, emphasizing the need for continuous learning and adaptation in a rapidly changing technological landscape. For example, DigCompEdu identifies six key areas of digital competence, including professional engagement, digital resources, and empowering learners through technology (Redecker & Punie, 2017). By aligning with such frameworks, educational institutions can create a supportive environment that encourages innovation and fosters pedagogical readiness.

Despite these efforts, disparities in pedagogical readiness persist across different educational contexts. While some educators demonstrate a high level of digital fluency, others struggle due to systemic barriers, such as inadequate infrastructure, resistance to change, and a lack of institutional encouragement (Tondeur et al., 2017). For instance, in developing countries, limited access to technology and insufficient funding often hinders the adoption of digital teaching methods (Sánchez et al., 2011). Even in developed nations, disparities exist between urban and rural schools, with the latter often facing challenges related to connectivity and resource availability (Zhao & Frank, 2003). Addressing these disparities requires a holistic approach that considers individual teacher needs, school-wide policies, and broader systemic support mechanisms (Cheah et al., 2023).

Institutional support and leadership [2] play a crucial role in fostering pedagogical readiness. School leaders who prioritize digital innovation and provide ongoing professional development opportunities create an environment where teachers feel empowered to experiment with new technologies (Davis & Hennessy, 2020). Transformational leadership, characterized by a shared vision and collaborative decision-making, has been shown to positively influence teachers' willingness to innovate (Fairbanks & Sheppard, 2023). Additionally, policies that promote equitable access to technology and encourage risk-taking can help bridge the gap between early adopters and reluctant educators (Kimmons et al., 2020).

Again, because pedagogical preparedness is a complicated phenomenon, a mixed-methods approach is necessary to fully capture its complexities. Teachers' experiences, difficulties, and perspectives can be illuminated by qualitative data, while quantitative data can reveal the prevalence of digital competence and the factors driving technology adoption (Creswell & Plano Clark, 2017). These methods can be combined to give academics a thorough grasp of the factors that support and hinder digital innovation in education (Ivankova et al., 2022). This study uses this methodology to investigate pedagogical preparedness in various learning contexts in order to add to the conversation about digital innovation and offer practical suggestions for teachers, legislators, and school administrators.

**Research Problem Statement**

Digital technology's explosive growth in education has produced previously unheard-of chances to improve instruction and student learning. But even if digital learning environments are becoming more and more important, instructors' willingness to adopt digital innovation varies and is not uniform across educational contexts. Although many educators agree that technology may improve student engagement, provide individualized learning experiences, and give access to global resources, they frequently face major obstacles that prevent successful integration (Al-Furaih & Al-Awidi, 2018). Adapting educational approaches to effectively use digital resources is one of these issues, along with a lack of institutional support and inadequate infrastructure (Ertmer et al., 2012).

One of the most pressing issues is the disparity in access to resources and training. Schools in under-resourced areas often lack the necessary technological infrastructure, such as reliable internet connectivity, up-to-date devices, and software, which creates a barrier to digital adoption [3] (Sánchez et al., 2011). Even in well-resourced institutions, teachers may struggle to integrate technology due to insufficient professional development opportunities or a lack of alignment between training programs and their specific needs (Cheah et al., 2023). Without adequate support, educators are left to navigate the complexities of digital innovation on their own, leading to frustration and resistance.

Resistance to change is another critical barrier to pedagogical readiness. Many teachers, particularly those with years of experience in traditional teaching methods, may view technology as a disruptive force rather than a tool for enhancing learning (Howard & Mozejko, 2015). This resistance is often fueled by a lack of confidence in using digital tools, fear of failure, or concerns about the time and effort required to master new technologies (Scherer et al., 2019). Additionally, some educators perceive technology as a threat to their professional autonomy, fearing that it may replace their role in the classroom or undermine their teaching expertise (Ertmer et al., 2012).

Moreover, educators face a special problem due to the quick speed of technology innovation. Teachers must constantly update their knowledge and modify their instructional practices in order to stay relevant as new technologies and platforms are developed (Koehler & Mishra, 2009). Professional development must be proactive in this dynamic environment, and institutions must have policies that support experimentation and innovation (Kimmons et al., 2020). Nevertheless, many schools lack the leadership and organizational structures required to assist teachers in this continuous process of change.

Given these challenges, there is an urgent need to assess the key factors influencing teachers' readiness for digital innovation and identify strategies to bridge the gap between technological advancements and classroom implementation. This study aims to address this gap by exploring the interplay between individual, institutional, and systemic factors that shape pedagogical readiness. This study looks at the experiences, viewpoints, and difficulties faced by teachers in various settings in an effort to offer useful information to legislators, school administrators, and teacher preparation programs. The ultimate objective is to enable teachers to fully utilize digital tools, resulting in more effective, inclusive, and engaging learning environments for every student.

**Objectives and Hypotheses**

The primary objectives of this study are:

1. To evaluate the current state of teachers' pedagogical readiness for digital innovation.
2. To examine the role of professional development in enhancing digital competence.
3. To explore the impact of institutional support on technology integration.

Based on these objectives, the study hypothesizes that:

1. Teachers with prior exposure to structured digital training programs demonstrate higher levels of digital pedagogical readiness.
2. Institutional support positively influences teachers' willingness and ability to integrate technology into their teaching.
3. Teachers' beliefs about technology play a crucial role in shaping their readiness for digital innovation.

This study employs a mixed-methods approach to provide a comprehensive assessment of pedagogical readiness for digital innovation. Quantitative data is collected through surveys measuring teachers' digital competence, perceived barriers, and institutional support (Kimmons et al., 2020). Qualitative data is obtained through semi-structured interviews to gain deeper insights into teachers' experiences and challenges in adopting digital tools (Cheah et al., 2023). Data analysis involves statistical techniques to identify trends and thematic analysis to interpret qualitative responses. The findings aim to inform policymakers and educational institutions on strategies to enhance digital readiness among educators.

# Literature Review

**Discussion of Previous Research Related to the Topic**

Previous research on digital innovation in education has emphasized the critical role of pedagogical readiness in determining successful technology integration. Studies by Ertmer et al. (2012) and Tondeur et al. (2017) have established that teachers' beliefs significantly influence their ability to implement digital tools effectively in classrooms. Teachers who perceive technology as beneficial for student learning are more likely to adopt innovative digital practices, whereas those who are skeptical may resist its implementation (Howard & Mozejko, 2015).

The Technological Pedagogical Content Knowledge (TPACK) framework proposed by Koehler and Mishra (2009) has been widely adopted in analyzing the intersection of digital skills, pedagogy, and subject knowledge. The framework emphasizes that effective technology integration requires a dynamic interplay among three core components: technological knowledge (TK), pedagogical knowledge (PK), and content knowledge (CK). Recent studies have expanded on TPACK by incorporating contextual factors, such as institutional policies and professional development opportunities, which significantly impact a teacher's ability to integrate technology effectively (Kimmons et al., 2020).

Institutional support has been identified as a crucial factor in fostering digital readiness among educators. Research by Al-Furaih and Al-Awidi (2018) highlights that teachers in institutions with structured digital training programs demonstrate higher confidence and competence in using technology for instructional purposes. Additionally, professional development opportunities tailored to technology integration have been found to enhance teachers' digital fluency, allowing them to apply innovative teaching strategies more effectively (Scherer et al., 2019). Despite this, many professional development programs focus primarily on technical skills rather than pedagogical strategies, leading to gaps in effective technology use in the classroom (Tondeur et al., 2017).

**Identification of Gaps in Existing Studies**

While existing literature extensively covers teacher beliefs and technological competence, there remains a gap in understanding how institutional policies and professional development programs directly impact pedagogical readiness (Cheah et al., 2023). Institutional factors such as administrative support, funding for technology resources, and alignment with national education policies play a significant role in shaping teachers' ability to integrate digital tools effectively. However, limited empirical studies have explored the direct correlation between these institutional factors and teachers' readiness for digital innovation.

Again, studies focusing on cross-cultural comparisons of digital readiness are limited. The majority of research has been conducted in technologically advanced countries, leaving a gap in understanding how educators in developing regions navigate digital transformation within resource-constrained environments (Althubyani, 2023). This lack of diversity in research samples limits the generalizability of findings and underscores the need for comparative studies that examine digital readiness across different educational contexts.

Another critical gap pertains to the long-term sustainability of digital competence among educators. While numerous studies document the immediate effects of professional development programs, fewer have investigated how teachers' digital competencies evolve over time with ongoing training and institutional support (Kimmons et al., 2020). Given the rapid pace of technological advancements, there is a need for longitudinal studies that assess how teachers continuously adapt to emerging digital tools and pedagogical innovations.

**Theoretical Framework Supporting the Study**

This research is anchored in the Technological Pedagogical Content Knowledge **(TPACK)** framework, as conceptualized by Koehler and Mishra (2009). TPACK offers a robust lens for examining how educators effectively integrate technology into their teaching practices. At its core, TPACK posits that effective technology integration necessitates a confluence of three key knowledge domains:

**Technological Knowledge (TK):** This encompasses a deep understanding of digital tools and their functionalities. Teachers with strong Technological Knowledge (TK) possess proficiency in utilizing various software applications, online learning platforms, and emerging technologies such as artificial intelligence (AI), augmented reality (AR), virtual reality (VR), and adaptive learning systems in education. TK extends beyond basic digital literacy to include knowledge of digital security, data privacy, and ethical considerations in technology use. Teachers proficient in TK can critically evaluate which tools best align with their pedagogical goals, how to seamlessly integrate them into lessons, and how to troubleshoot technical issues that may arise during instruction. Additionally, TK involves staying updated on technological trends and being adaptable to emerging innovations that can enhance learning experiences.

**Pedagogical Knowledge (PK):** This domain encompasses a comprehensive understanding of effective teaching methodologies and instructional strategies. Teachers with strong PK have knowledge of varied learning theories (e.g., constructivism, behaviorism, connectivism), differentiated instruction techniques, and formative and summative assessment strategies. PK enables educators to design learner-centered, inquiry-based, and interactive learning environments that foster engagement and critical thinking. Furthermore, PK involves expertise in classroom management techniques, student motivation strategies, and the ability to scaffold learning experiences to accommodate diverse student needs. Teachers proficient in PK can adapt their instructional approaches based on the context, student demographics, and subject requirements, ensuring an inclusive and equitable learning experience.

**Content Knowledge (CK):** This refers to a deep and nuanced understanding of the subject matter being taught. CK goes beyond factual knowledge to include an understanding of the underlying concepts, principles, and interconnections within a discipline. Teachers with strong CK can explain complex ideas in an accessible manner, draw cross-disciplinary connections, and apply real-world examples to make content relevant to students. Additionally, CK involves staying current with advancements in the field, engaging in continuous professional learning, and collaborating with subject-matter experts to ensure content accuracy and relevance. Strong CK is essential for fostering higher-order thinking skills in students, encouraging critical analysis, problem-solving, and creativity within the subject domain.

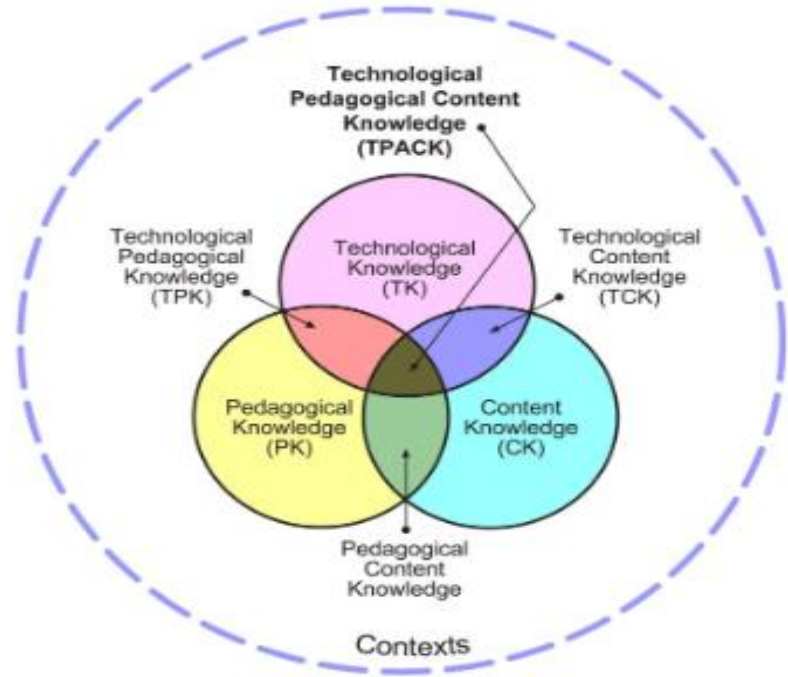

TPACK emphasizes that these three knowledge domains are not independent but rather interconnected and interdependent. Effective technology integration requires teachers to understand how technology can be used to support and enhance their pedagogical practices and how these practices can be adapted to effectively convey the subject matter.

Furthermore, this study incorporates the **PICRAT model** (Kimmons et al., 2020) to further refine the analysis. The PICRAT model provides a valuable framework for evaluating the quality of technology integration in the classroom. It examines technology use along two key dimensions:

> **PIC (Passive, Interactive, Creative):** This dimension assesses the level of student engagement facilitated by technology. It explores whether technology is used primarily for passive information delivery or if it actively engages students in higher-order thinking skills such as problem-solving, critical thinking, and creativity.
>
> **RAT (Replacement, Amplification, Transformation):** This dimension examines how technology is used to enhance teaching and learning practices. It assesses whether technology is used to simply replace traditional methods (e.g., using a projector instead of a chalkboard), amplify existing practices (e.g., using online resources to supplement textbook readings), or fundamentally transform teaching and learning (e.g., creating interactive simulations, using data analytics to personalize learning).

Through the integration of the TPACK and PICRAT models, this research endeavors to offer a thorough comprehension of the elements that contribute to successful classroom technology integration. The findings will have a big impact on professional development, teacher education, and creating technologically advanced learning environments that help students succeed.

# Methodology

## Research Design

This study adopts a mixed-methods approach, integrating both quantitative and qualitative data collection techniques. The mixed-methods design allows for a comprehensive analysis of teachers' pedagogical readiness by combining numerical data with in-depth insights into their experiences (Creswell & Plano Clark, 2017). The rationale behind this approach is to triangulate findings, ensuring validity by cross-verifying data from multiple sources and providing a nuanced understanding of the factors influencing digital innovation adoption. A convergent parallel design was used, in which both quantitative and qualitative data were collected simultaneously, analyzed separately, and then merged to enhance the reliability of the findings.

## Data Collection Methods

To ensure a comprehensive understanding of teachers' pedagogical readiness for implementing innovative activities, multiple data collection methods were utilized. These methods provided both quantitative and qualitative insights, capturing various dimensions of technological competence, institutional support, and real-world application of digital tools in teaching.

> **Surveys**: Surveys were employed as a primary data collection tool to quantitatively assess teachers' technological competence, beliefs, and the institutional structures that support digital integration. A structured questionnaire was developed, incorporating both closed-ended and open-ended questions. The closed-ended questions utilized a five-point Likert scale ranging from "Strongly Disagree" to "Strongly Agree" to measure teachers' confidence in using digital tools, their perceptions of institutional readiness, and their attitudes toward pedagogical innovation. Additionally, open-ended questions allowed participants to elaborate on their experiences, challenges, and expectations, providing deeper insights into factors influencing their digital readiness.
>
> **Interviews**: Interviews were conducted with selected teachers to supplement the survey data with in-depth qualitative insights. A semi-structured interview format was used, allowing for a guided conversation while also providing flexibility for participants to discuss relevant personal experiences. The interviews explored key themes such as the perceived benefits and barriers of integrating technology into teaching, institutional support mechanisms, and teachers' adaptability to new digital tools. The responses gathered from these interviews enriched the research by offering nuanced perspectives that could not be captured through survey responses alone.
>
> **Observation**: Observations were carried out in real classroom environments to assess the actual application of digital tools in pedagogical practices. The observations focused on multiple aspects, including how teachers interact with digital platforms, the level of student engagement with technology, and the extent to which institutional support facilitates smooth technological integration. Specific attention was given to how teachers adapted their instructional strategies when incorporating digital resources and whether challenges such as technical issues or lack of digital literacy impacted lesson delivery. By analyzing classroom interactions, the study was able to validate self-reported data from surveys and interviews, ensuring a more accurate representation of teachers' digital readiness.

**Document analysis**: Document analysis was conducted to examine existing institutional policies, professional development programs, and curriculum guidelines related to digital integration. Official documents, such as professional training records, educational policies, and strategic plans, were analyzed to understand the structural framework that supports or hinders teachers' readiness for pedagogical innovation. Reviewing these documents provided insights into the extent to which schools prioritize digital literacy, the availability of professional development opportunities, and whether existing policies align with the practical realities observed in classrooms. This method allowed for the identification of gaps [4] between institutional objectives and actual implementation in educational settings.

**Sampling Techniques and Participant Details**

A stratified random sampling technique was employed to ensure representation across different educational levels and institutional types. The stratification was based on school type (primary, secondary, and higher education) and teaching experience levels to account for variations in digital readiness. The final sample consisted of 200 teachers, distributed as follows:

| Category | Number of Participants |
|---|---|
| Primary Schools | 75 |
| Secondary Schools | 75 |
| Higher Education | 50 |

To further contextualize the findings, additional demographic data were collected. These included years of teaching experience, which provided insights into whether more experienced educators adapted differently to digital innovations compared to their early-career counterparts. The study also gathered information on prior exposure to digital training, which helped assess whether professional development opportunities played a role in shaping digital readiness. Furthermore, accessibility to digital resources was examined, as disparities in infrastructure and technological support could significantly influence the effectiveness of digital implementation in different educational settings.

**Data Analysis Methods**

To derive meaningful insights from the collected data, a combination of statistical and thematic analysis methods was employed. This dual approach enabled a comprehensive evaluation of teachers' pedagogical readiness by integrating quantitative assessments with qualitative interpretations. The analysis focused on identifying key determinants of digital readiness, exploring relationships between influencing factors, and ensuring data reliability through triangulation.

Quantitative data analysis was conducted to systematically process and interpret the survey responses. Descriptive statistics, including mean, standard deviation, and frequency distribution, were used to summarize participants' technological competence, institutional support, and self-perceptions regarding digital integration. These statistical measures provided an overall understanding of the dataset's central tendencies and variations. Regression analysis was performed to examine the relationships between teacher beliefs, institutional support structures, and digital readiness, offering insights into how these factors interact to influence educators' ability to integrate digital tools effectively (Scherer et al., 2019). Furthermore, chi-square tests were applied to determine whether significant differences existed among demographic groups, allowing for the identification of disparities in digital readiness based on variables such as age, years of experience, and institutional setting.

Qualitative data analysis involved an in-depth exploration of interview transcripts and classroom observation notes to extract meaningful patterns and insights. Thematic analysis was conducted to identify recurring themes related to challenges, attitudes, and best practices in digital adoption. This process allowed for the classification of key barriers and facilitators that impact teachers' engagement with technology (Kimmons et al., 2020). Coding techniques, including open coding and axial coding, were applied to systematically categorize emerging themes and establish connections between them. To enhance the validity of findings, data triangulation was employed by cross-referencing insights from interviews, observations, and document analysis, ensuring consistency and reducing bias.

This study offers a thorough, empirically supported evaluation of pedagogical preparedness for digital innovation by combining various analytical techniques. The results aid in the creation of plans to improve teachers' technology competence, remove institutional obstacles, and improve professional development programs in order to facilitate the digital revolution of education.

## Results & Discussion

### Key Findings

### Quantitative Results

> **Digital Competence**: 68% of teachers in the study reported feeling confident in using digital tools, indicating a strong sense of self-assurance in their ability to navigate technology. However, only 45% of these educators integrated digital tools regularly into their lessons (Kimmons et al., 2020). This significant gap between teachers' confidence and their actual implementation of digital tools suggests that despite their readiness, barriers may exist that hinder the consistent use of technology in classrooms. These barriers may include a lack of institutional support, insufficient training, limited access to updated resources, or even time constraints. The discrepancy highlights the need for additional strategies to bridge the gap between teachers' perceived competence and their practical application of digital tools in their pedagogical practices.
>
> **Institutional Support**: Schools that offered structured and ongoing professional development programs experienced a 30% higher adoption rate of digital tools compared to those that lacked such programs (Ertmer et al., 2012). This finding underscores the critical role that institutional support plays in enhancing teachers' digital competence. Continuous professional development provides educators with the necessary knowledge and skills to effectively incorporate new technologies into their teaching. The support of school leadership, alongside a structured framework for training, ensures that teachers are not only confident but also equipped to leverage digital tools in a meaningful way. Without such support, even teachers with strong initial confidence may struggle to implement technology effectively in the classroom.

| Factor | Percentage (%) |
|---|---|
| Confident in Tech Use | 68% |
| Regular Tech Integration | 45% |
| Strong Institutional Support | 30% higher adoption rate |

**Qualitative Results**

Teachers overwhelmingly cited the lack of professional development as a major barrier to digital integration (Howard & Mozejko, 2015). Many educators reported that available training sessions were either too infrequent or not tailored to their specific instructional needs, leaving them feeling ill-equipped to integrate digital tools effectively into their teaching. Without targeted and continuous professional learning opportunities, teachers struggled to bridge the gap between digital competence and practical classroom application.

Resistance to change was particularly evident in institutions with rigid administrative structures, where digital innovation was perceived as an additional burden rather than an essential pedagogical shift (Tondeur et al., 2017). In these settings, teachers often expressed frustration with the lack of institutional encouragement and limited incentives to adopt digital tools. Some educators noted that existing school policies prioritized traditional teaching methods, creating an environment where experimentation with technology was discouraged or undervalued. This institutional inertia not only stifled innovation but also contributed to a culture of apprehension toward digital transformation.

In addition to professional and institutional barriers, technical challenges further hindered effective digital adoption. Some educators pointed out issues such as unreliable internet connectivity, outdated devices, and a general lack of IT support within schools. These infrastructural deficiencies made it difficult for teachers to consistently integrate digital tools into their lessons, even when they were willing to do so. Many teachers emphasized the need for improved access to up-to-date technology and dedicated technical assistance to troubleshoot digital challenges, ensuring smoother implementation of digital teaching strategies.

**Interpretation of Results**

The findings indicate that while teachers acknowledge the importance of digital innovation, many faces persistent challenges related to training, institutional policies, and access to resources. The discrepancy between self-reported confidence (68%) and actual integration (45%) suggests that psychological and structural barriers play a significant role in digital adoption. Teachers require not just technical skills but also pedagogical strategies to integrate technology effectively into their curriculum.

The study's results align with previous research highlighting the need for comprehensive, ongoing professional development programs (Koehler & Mishra, 2009). Effective technology integration is not a one-time event but an evolving process that requires consistent support, mentorship, and access to updated digital tools.

**Comparison with Existing Literature**

Findings from this study align with prior research emphasizing that teacher beliefs and institutional support play a pivotal role in the adoption of digital technologies in education. Ertmer et al. (2012) and Scherer et al. (2019) have highlighted that teachers' self-efficacy and confidence in using technology significantly influence their willingness to integrate digital tools into their teaching practices. This study reinforces those conclusions while extending them by demonstrating that structured professional development programs are directly linked to increased digital readiness. The evidence gathered suggests that training initiatives tailored to educators' needs not only improve their technological proficiency but also foster a more positive attitude toward digital transformation.

While previous research has consistently established that teachers with favorable attitudes toward technology are more likely to incorporate it into their pedagogy, this study expands on that perspective by illustrating the equally critical role of institutional support. Factors such as administrative encouragement, dedicated IT support, and a well-defined digital strategy contribute significantly to the success of technology adoption in educational settings. This finding is in line with Scherer et al. (2019), who emphasized the necessity of a supportive ecosystem for sustainable digital integration.

The study also observed a level of resistance to change among certain educators, a trend documented by Howard and Mozejko (2015). Their research found that schools with rigid structures, bureaucratic constraints, and lack of incentives for innovation tend to discourage the adoption of new teaching methodologies. This study corroborates that finding, demonstrating that teachers operating within such environments often express hesitancy toward technology integration, citing concerns about workload, lack of training, and limited administrative backing.

Furthermore, compared to existing studies that predominantly focus on individual teacher competence, this research underscores the systemic nature of digital adoption. While personal skills and attitudes are crucial, this study highlights the broader importance of infrastructure, leadership engagement, and policy interventions in shaping an institution's digital transformation. Schools with clear technology implementation policies, accessible resources, and leadership that prioritizes digital literacy tend to have higher overall digital readiness among educators.

**Implications for Practice, Policy, and Future Research**

>  **For Practice**: Schools should implement structured training programs designed to enhance digital readiness among educators (Cheah et al., 2023). These programs should extend beyond basic IT skills to encompass pedagogical strategies that enable teachers to integrate technology seamlessly into their instructional practices. Effective professional development should include hands-on workshops, peer collaboration, and real-world application scenarios to ensure that teachers can translate their digital knowledge into meaningful classroom experiences. Additionally, ongoing support and mentorship should be provided to help educators navigate technological challenges and continuously refine their digital teaching strategies.

>  **For Policy**: Educational institutions should revise existing policies to establish stronger support systems for teachers adopting digital technologies (Kimmons et al., 2020). This includes increasing access to essential IT resources, such as updated software, high-speed internet, and interactive learning platforms. Schools should also consider appointing digital coaches or technology mentors who can provide individualized guidance to educators, helping them integrate digital tools effectively into their curriculum.

>  **For Future Research**: Further studies should explore the longitudinal impacts of professional development programs on digital competence among educators (Creswell & Plano Clark, 2017). While short-term studies provide valuable insights into immediate changes in teachers' skills and attitudes, there is a need to examine how sustained engagement with digital training influences long-term technology adoption. Future research should also investigate factors that contribute to the retention and application of digital skills over time, including the role of continuous professional learning communities, institutional culture, and evolving technological advancements in shaping educators' digital practices.

## Conclusion

This study has provided a comprehensive examination of pedagogical readiness for digital innovation, shedding light on both the opportunities and challenges associated with integrating technology into teaching practices. The findings reveal that while many educators recognize the importance of digital tools in enhancing learning outcomes, significant barriers hinder their effective adoption. Key findings include:

>  **Digital Competence**: Many teachers possess a foundational level of technological proficiency, enabling them to use basic tools such as presentation software and learning management systems. According to Simmons et al. (2020), there is frequently a deficiency in the advanced abilities necessary for innovative digital pedagogy, such as developing collaborative online learning environments, utilizing data analytics, and producing interactive content. Given this disparity, specific training programs that cover the pedagogical as well as technical facets of digital integration are important.

**Institutional Support**: Schools with robust professional development programs and a supportive infrastructure report higher levels of digital adoption among teachers (Ertmer et al., 2012). Institutional support, including access to up-to-date technology, funding for training, and policies that encourage experimentation, plays a critical role in fostering pedagogical readiness. Conversely, schools with limited resources or inadequate leadership often struggle to implement digital innovation effectively (Davis & Hennessy, 2020).

**Teacher Beliefs**: Positive attitudes toward technology significantly influence the likelihood of its integration into teaching practices. Educators who view digital tools as valuable for enhancing student engagement and learning outcomes are more likely to adopt them (Howard & Mozejko, 2015). However, resistance remains a persistent obstacle, particularly among teachers who perceive technology as a threat to traditional teaching methods or who lack confidence in their ability to use it effectively (Ertmer et al., 2012).

**Limitations of the Study**

While this study contributes valuable insights into pedagogical readiness for digital innovation, it is not without limitations. These limitations highlight areas that require further investigation to ensure a more comprehensive understanding of the factors influencing educators' ability to integrate digital tools effectively into their teaching practices:

**Sample Size and Scope**: The study was conducted with a limited number of teachers across specific educational institutions, which may not fully capture the diversity of experiences and challenges faced by educators in different contexts (Cheah et al., 2023). Future studies should aim to include a broader and more representative sample to enhance the generalizability of the findings.

**Self-Reported Data**: Much of the data was collected through surveys and interviews, which are inherently subject to personal biases and social desirability effects (Scherer et al., 2019). For instance, teachers may overstate their digital competence or underreport their resistance to technology due to perceived expectations. Triangulating self-reported data with observational or performance-based measures could provide a more accurate assessment of pedagogical readiness.

**Short-Term Focus**: This study provides a snapshot of pedagogical readiness at a single point in time, offering limited insights into how digital competencies evolve over time (Tondeur et al., 2017). Longitudinal studies are needed to track changes in teachers' skills, attitudes, and practices as they engage with ongoing professional development and adapt to emerging technologies.

**Recommendations for Future Research**

To build upon the findings of this study, future research should consider the following directions to further enhance the understanding of pedagogical readiness for digital innovation and address the existing gaps in literature

First, longitudinal studies should be conducted to examine changes in teachers' digital competencies over time can provide valuable insights into the long-term impact of professional development programs and institutional policies (Creswell & Plano Clark, 2017). Such studies could also identify critical junctures where additional support or intervention is needed to sustain digital innovation.

Secondly, cross-cultural comparisons should be done by investigating how pedagogical readiness varies across different educational systems and cultural contexts can help identify best practices and contextual factors that influence technology adoption (Al-Furaih & Al-Awidi, 2018). For example, comparing high-resource and low-resource settings may reveal strategies for overcoming systemic barriers to digital integration.

Again, intervention-based research should be embarked on by evaluating the effectiveness of targeted training programs in improving digital adoption rates among educators is essential for developing evidence-based practices (Koehler & Mishra, 2009). Future studies could explore the impact of specific interventions, such as mentorship programs, collaborative learning communities, or gamified training modules, on teachers' digital readiness.

Moreover, focusing on equity and inclusion should be considered. Future research should also examine how digital innovation can be leveraged to address educational inequities and support marginalized learners. This includes exploring the role of culturally responsive pedagogy and inclusive design in digital teaching practices (Cheah et al., 2023).

## Endnotes

[1] Defining Pedagogical Readiness: Pedagogical readiness encompasses not only technical skills but also the instructional strategies and mindset required to integrate digital tools effectively into teaching practices (Kimmons et al., 2020). It involves a shift from traditional, teacher-centered approaches to more student-centered, interactive, and collaborative learning environments.

[2] The Role of Leadership: Strong institutional leadership is critical for fostering a culture of innovation and encouraging educators to embrace digital transformation (Tondeur et al., 2017). Leaders who prioritize digital readiness, allocate resources for professional development, and model innovative practices can inspire confidence and motivation among teachers.

[3] Challenges in Digital Adoption: Resistance to technology integration is often rooted in fears of job displacement, concerns about increased workload, and a lack of adequate training (Howard & Mozejko, 2015). Addressing these challenges requires targeted policy interventions, such as providing incentives for technology use, reducing administrative burdens, and offering ongoing support for skill development.

[4] Professional Development Gaps: Many training programs focus on technical proficiency but fail to address the pedagogical shifts required for meaningful technology integration (Ertmer et al., 2012). Effective professional development should emphasize the alignment of technology with curriculum goals, the creation of engaging and inclusive learning experiences, and the use of data to inform instructional decisions.

## Conflict of Interest Statement

The author declares that there is no conflict of interest.